\documentclass[aps,preprint,floatfix,nofootinbib,showpacs]{revtex4-1}
\pdfoutput=1
\usepackage{graphicx,color}
\usepackage{hyperref}
\usepackage{amsmath}
\usepackage{amsfonts}
\usepackage{amssymb}
\usepackage{setspace}

\begin{document}
\title{Interpretation of excess in $H \to Z \gamma$ using a light axion-like
  particle}

\author{Kingman Cheung$^{a,b,c}$ and C.J.~Ouseph$^{a,b}$ }
\affiliation{$^a$ Department of Physics, National Tsing Hua University,
  Hsinchu 300, Taiwan}
\affiliation{$^b$ Center for Theory and Computation, National Tsing Hua
  University, Hsinchu 300, Taiwan}
\affiliation{$^c$ Division of Quantum Phases and Devices, School of Physics,
   Konkuk University, Seoul 143-701, Republic of Korea}

\email{cheung@phys.nthu.edu.tw, ouseph444@gmail.com}

\date{\today}

\begin{abstract}
  We interpret the recent excess in a rare decay of the Higgs boson,
  $H\to Z\gamma$, using a light axion-like particle (ALP) in the mass
  range $0.05 -  0.1$ GeV.
  The dominant decay of such a light ALP is into a pair of collimated 
  photons,   whose decay is required to happen before reaching the
  ECAL detector, such that it mimics a single photon in the detector.
  It can explain the excess with a coupling 
  $C^{\rm eff}_{aZH} / \Lambda \sim 4 \times 10^{-5}\;{\rm GeV}^{-1}$,
  while the decay of the ALP before reaching the ECAL requires
  the diphoton coupling
  $C^{\rm eff}_{\gamma\gamma}/ \Lambda \ge 0.35 \,{\rm TeV}^{-1}
  (0.1\,{\rm GeV}/m_a)^2$.
  A potential test would be the rare decay of the $Z$ boson
  $Z \to a H^* \to a (b \bar b)$ at the Tera-$Z$ option of
  the future FCC and CEPC. However, it has a branching ratio of
  only $O(10^{-12})$, and thus barely testable.
  The production cross section for $pp \to Z^* \to a H$ via 
  the same coupling $C^{\rm eff}_{aZH} / \Lambda$ at the LHC
  is too small for detection.
\end{abstract}

\maketitle

\section{Introduction}
Since the discovery of the Higgs boson in 2012
\cite{ATLAS:2012yve,CMS:2012qbp},
all the gauge couplings and the
third-generation Yukawa couplings are shown to be
consistent with the standard model (SM) Higgs boson (see the most recent fits
\cite{Heo:2024cif}), including the loop-induced $Hgg$ and $H\gamma\gamma$
couplings.  The $H \to Z \gamma$ is one of the most anticipated measurements
of the Higgs physics. Recently, an evidence of such a rare decay
$H \to Z \gamma$ was jointly reported by ATLAS and CMS \cite{ATLAS:2023yqk}.
The search showed an observed significance of $3.4$ standard deviations
from the null hypothesis. The measured branching ratio of $H\to Z\gamma$:
\begin{equation}
  B(H \to Z \gamma )_{\rm measured} = (3.4 \pm 1.1) \times 10^{-3} \;.
\end{equation}
The SM prediction for the branching ratio of  $H\to Z\gamma$
is \cite{Djouadi:1997yw}
\begin{equation}
  B(H \to Z\gamma)_{\rm sm}  = (1.5\pm 0.1) \times 10^{-3} \;.
  \end{equation} 
It is clear that the measurement showed an excess of $1.9\, \sigma$
\cite{ATLAS:2023yqk}.

\begin{figure}[th!]
  \centering
  \includegraphics[width=5in]{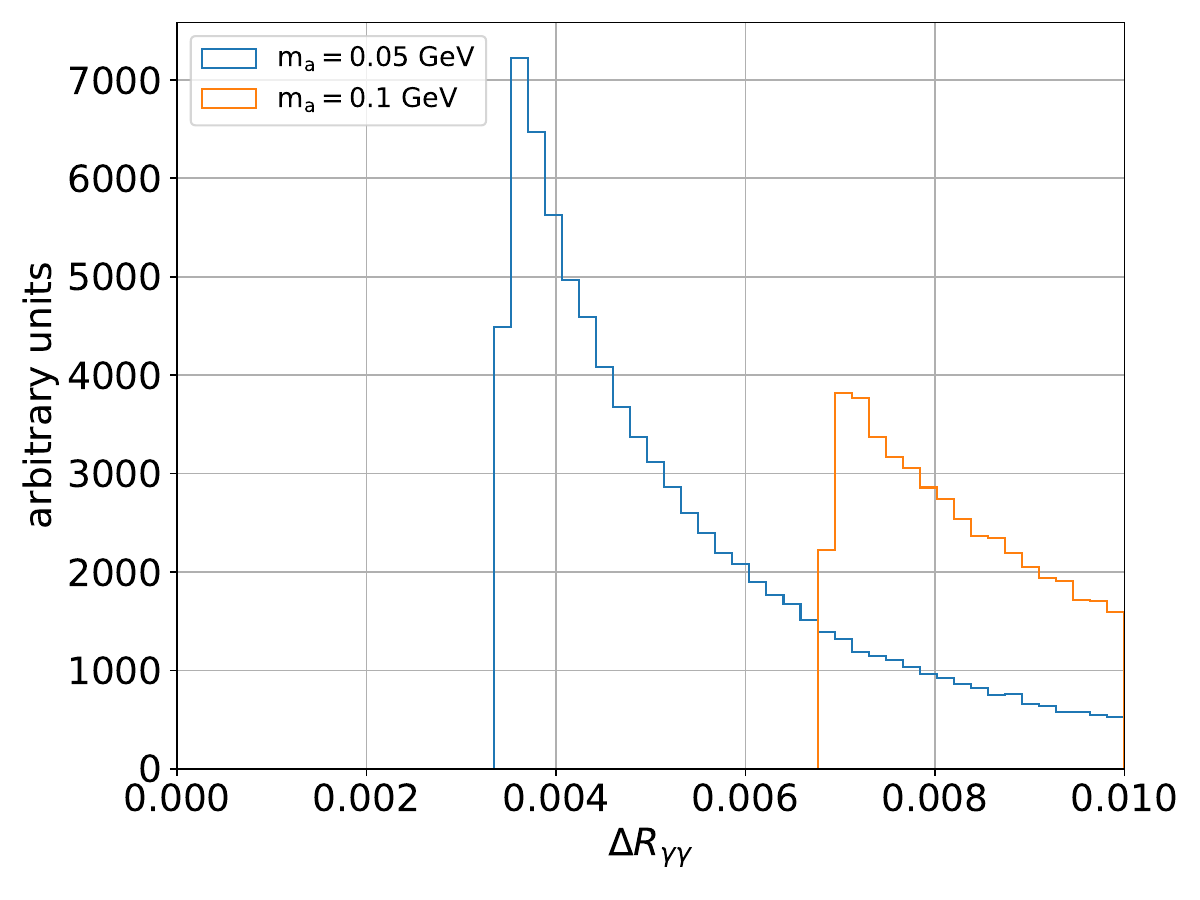}
  \caption{\small \label{deltar}
    Distributions of $\Delta R_{\gamma\gamma}$ between the photon pair
    produced for $m_a =0.05$ GeV and $0.1$ GeV
    in the decay $H \to Z a \to (l^+ l^-)\,(\gamma\gamma)$.
    It is clear the opening angle between the photon pair is very small.
   }
  \end{figure}

 In various well-founded extensions of the Standard Model (SM), there
 is a common occurrence of newly discovered pseudoscalar particles
 possessing masses lower than the electroweak scale. These particles
 serve various purposes, such as addressing the strong CP problem like
 axions~\cite{Peccei:1977hh,Peccei:1977ur,Weinberg:1977ma,Wilczek:1977pj,Kim:1979if,Shifman:1979if,Zhitnitsky:1980tq,Dine:1981rt}
 or acting as pseudoscalar mediators facilitating interaction between
 dark or hidden sectors and the SM~\cite{Dolan:2014ska}.
Although it may be too early to conclude that new physics exists in the
rare decay $H \to Z\gamma$, we speculate the interpretation of the excess
using a very light axion-like particle (ALP) of mass
$m_a = 0.05 -  0.1$ GeV.
For such a light ALP the dominant decay of the ALP is $a\to \gamma\gamma$.
Since the ALP is produced in the decay of the Higgs boson, we expect the
transverse momentum $p_{T_a}$ of the ALP is of order
  $ (m_H/2) ( 1 - m_Z^2 /m_H^2) \simeq m_H /4$,
  taking into account the massive $Z$ boson.
It is well known that
the opening angle $\Delta R$ between the decay products of the ALP is then
\[
\Delta R \sim \frac{ 2 m_a }{ p_{T_a} } \approx
(3  - 7 ) \times 10^{-3} \;,
\]
for $m_a = 0.05 - 0.1 $ GeV.
We show the $\Delta R$ distributions for $m_a = 0.05$ GeV and 0.1 GeV in
Fig.~\ref{deltar}.
Both the ATLAS and CMS detectors cannot resolve the two photons in
such a small opening angle
  \cite{ATLAS:2018fzd,CMS:2020uim}.
In this case, both photons deposit their
energies in a single cell.
In order that it happens, the axion has to decay before reaching or inside
the ECAL detector. It is the coupling $C^{\rm eff}_{\gamma\gamma}/\Lambda$ in
Eq.~(\ref{eq:gamma}) that controls the decay length of the ALP.
The ECAL detector the ATLAS detector extends from the radius of 1.5 m to
2 m while that the CMS is slightly closer to the center. We therefore
require the decay length of the axion to be less than 1.5 m. 
When these conditions are met, the diphoton decay of the axion 
would be mistaken as a single photon, and thus mimics the
$H \to Z \gamma$ decay.

Details of the experimental event selections have been given in the CMS
and ATLAS publications \cite{CMS:2022ahq,ATLAS:2020qcv}.
In both detectors, photons are identified as ECAL energy clusters
not linked to the extrapolation of any charged particle trajectory
to the ECAL. Typical angular resolution of the ECAL is of order
$\Delta R \sim 10^{-2}$, which is given by the size of each cell,
for photon energies of order $50-100$ GeV
\cite{Aleksa:2018xmp,ATLASLiquidArgonCalorimeter:2005bfr}.
  It was demonstrated that taking advantage of a shower-shape analysis
  \cite{ATLAS:2012soa} (also 
emphasized in Ref.~\cite{Bauer:2017ris}) the ECAL detector is able to
distinguish a single photon from a pair of collimated photons
for $m_a \agt 0.1$ GeV.
That is the reason for the upper limit of $m_a = 0.1$ GeV
that we propose while the lower limit 0.05 GeV is given by the
existing limit on $C^{\rm eff}_{\gamma\gamma}/ \Lambda$.

The final state of $Z a \to (\ell^+ \ell^-) ( \gamma
\gamma)$ mimics $(Z\to \ell^+ \ell^-) \gamma$.
Taking the difference between the measurement
$B(H \to Z \gamma)_{\rm measured}$ and the SM prediction of
$B(H\to Z \gamma)_{\rm sm}$ is entirely due to $H \to Z a$, we obtain
\begin{equation}
  \label{1.9}
  B ( H \to Z a ) = (1.9 \pm 1.1) \times 10^{-3}  \;.
\end{equation}
In this work, we show that an effective coupling among $a$-$Z$-$H$,
$C^{\rm eff}_{aZH} / \Lambda \approx 4.4 \times 10^{-5}\;{\rm GeV}^{-1}$,
can explain the excess, without violating any other existing constraints.
We also show that this interpretation may be tested at the Tera-$Z$ option
of the FCC \cite{FCC:2018byv} and CEPC \cite{CEPCStudyGroup:2018ghi}.
On the other hand, the production cross section for 
$pp \to Z^* \to a H \to (\gamma\gamma) (b \bar b)$ via the same coupling of
$aZH$ at the LHC is negligible for detection.

A few other interpretations were also proposed
\cite{Barducci:2023zml,Boto:2023bpg,Das:2024tfe}.
Barducci et al. \cite{Barducci:2023zml} used extra chiral leptons with
hypercharge $Y$ and with scanning some choices of hypercharge $Y$ the
$H\to \gamma Z $ can be enhanced without increasing $H\to \gamma\gamma$.
Boto et al. \cite{Boto:2023bpg} used multiple charged scalar bosons $S^{+Q}_i$,
which couple to $H$ and $Z$, and with enhanced off-diagonal couplings,
the authors can increase $H \to \gamma Z$ without increasing
$H\to \gamma\gamma$.
Das et al. \cite{Das:2024tfe} made use of the triplet scalar field in the
context of Type II seesaw model and adjusted the couplings of the
singly- and doubly-charged scalars to achieve the enhancement of
$H \to Z\gamma$ without increasing $H\to \gamma\gamma$.

  There are numerous studies of collimated photons from axion or ALP decays
  in literature: see for example
  \cite{Draper:2012xt,Ellis:2012zp,Alonso-Alvarez:2023wni,Lane:2023eno}.

\section{Model}
We follow the notation of Ref.~\cite{Bauer:2018uxu}.
The interactions of the ALP $a$ with the SM particles start at
dimension-5~\cite{Georgi:1986df}:
\begin{eqnarray}
  {\cal L}^{D=5} &=& \frac{1}{2} (\partial_\mu a) (\partial^\mu a) 
  - \frac{1}{2} m_a^2 a^2
  + \sum_f \frac{c_{ff}}{2\Lambda} \partial^\mu a \,
  \bar f \gamma_\mu \gamma_5 f  \nonumber \\
  &+& g_S^2 \frac{C_{GG}}{\Lambda} a G^A_{\mu\nu} \tilde{G}^{\mu\nu,A}
  + g^2  \frac{C_{WW}}{\Lambda} a W^i_{\mu\nu} \tilde{W}^{\mu\nu,i}
  +g'^2 \frac{C_{BB}}{\Lambda} a B_{\mu\nu} \tilde{B}^{\mu\nu} \;,
  \label{eq4}
\end{eqnarray}
where $A = 1,....8$ is the $SU(3)$ color index, $i = 1,2,3$
is the $SU(2)$ index, and $g_S$, $g$ and $g'$ are the gauge couplings
of $SU(3)$, $SU(2)$ and $U(1)_Y$, respectively.
We set $C_{GG}=0$ to avoid the mixing of the ALP with
the QCD axion such that the strong CP problem would not come back.
After the $B_\mu$ and $W^3_{\mu}$ rotate into the physical $\gamma,Z$, the
ALP couples to $\gamma$ and $Z$ as
 \begin{eqnarray}
   {\cal L} &=& e^2 \frac{C_{\gamma\gamma} }{\Lambda} a F_{\mu\nu}
   \tilde{F}^{\mu\nu}
   +  \frac{2e^2}{s_w c_w} \frac{C_{\gamma Z} }{\Lambda} a F_{\mu\nu}
   \tilde{Z}^{\mu\nu}
   +  \frac{e^2}{s^2_w c^2_w} \frac{C_{ZZ} }{\Lambda} a Z_{\mu\nu}
   \tilde{Z}^{\mu\nu} \;, \label{eq:gamma}
  \end{eqnarray}
 where
 \[
  C_{\gamma\gamma} = C_{WW} + C_{BB}\;, \;\;\;
  C_{\gamma Z } = c_w^2 C_{WW} - s_w^2  C_{BB}\;, \;\;\;
  C_{ZZ} = c_w^4 C_{WW} + s_w^4 C_{BB}\;,
 \]
 and $s_w$ and $c_w$ are the sine and cosine of the weak mixing angle,
 respectively.
%
 In the considered mass range of the ALP $m_a \le 0.1$ GeV, the only decay
 modes are $e^+ e^-$ and $\gamma\gamma$, for which the $\gamma\gamma$ can
 entirely dominate for $O(1)$ coefficients. However, we set
 $C_{ff} = 0$ for simplicity. Even in this case, the $a \to e^+ e^-$ can
 be induced by a loop diagram, but it is largely suppressed. Therefore,
 the ALP so produced will decay entirely into a pair of photons.

 \begin{figure}[thb!]
  \centering
  \includegraphics[width=3.5in]{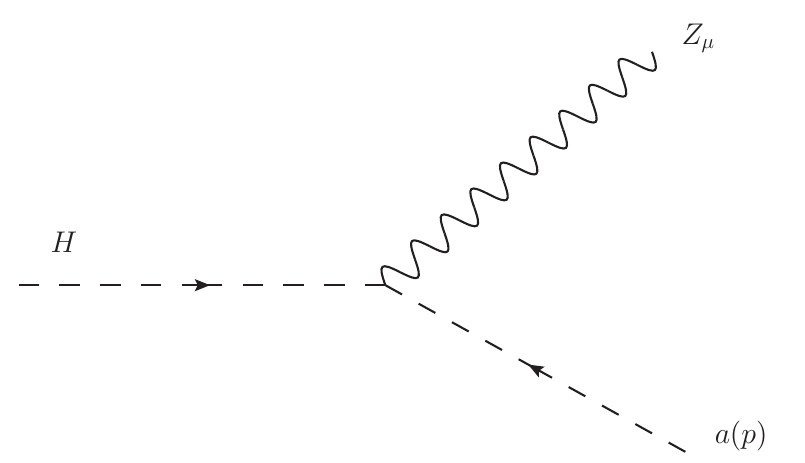}  
    {\Large $\frac{C^{\rm eff}_{aZH} }{\Lambda} \frac{g v}{c_w}\, p^\mu $ }
  \caption{\small \label{fey}
    Feynman rule for the vertex of $aZH$, which $p_\mu$ is the momentum of
    the {\it incoming axion}.}
\end{figure}
  
 Interactions with the Higgs boson start at dimension-6:
 \footnote{ The obvious dimension-5 operator
    $ (\partial^\mu a)( \phi^\dagger i D_\mu \phi + {\rm h.c.})$
    is reduced to the fermionic operators using the equations of
    motion and integration by parts, such that 
    this dimension-5 operator does not appear \cite{Bauer:2016ydr}.
    In another word, this dimension-5 operator cannot contribute
    to $H \to Za $ because there exists an equivalent basis in
    which this decay does not appear.
}
\begin{equation}
  \label{lag}
   {\cal L}^{D \ge 6}
   = \frac{C_{ah}}{\Lambda^2} (\partial_\mu a) (\partial^\mu a) \,
   \phi^\dagger \phi
   + \frac{C_{aZH}}{\Lambda^3} (\partial^\mu a) \,
   \left( \phi^\dagger i D_\mu \phi + {\rm h.c.} \right )\,
   \phi^\dagger \phi \;,
 \end{equation}
 where the covariant derivative is given by
 \[
 D_\mu = \partial_\mu + i \frac{g}{\sqrt{2}}
 \left( W^+_\mu \tau^+ + W^-_\mu \tau^- \right ) + i e Q A_\mu
 + i\frac{g}{c_w} (T_3 - s_w^2 Q ) Z_\mu \;,
 \]
 and $\tau^\pm$ are the $SU(2)$ raising and lowering operators,
 $T_3$ is the third component of the isospin, and $Q$ is the electric charge.
 It is easy to see that the first term in Eq.~(\ref{lag})
 induces the decay $H \to a a $ while
 the second term induces $ H \to Z a $. From dimensional analysis the
 amplitude for $H \to Z a$ is suppressed by one more order of the
 cutoff scale $\Lambda$ than $H\to a a$.
 However, as familiar to the Higgs low-energy theorems \cite{Kniehl:1995tn},
 in theories where a heavy new particle acquires most of its
 mass through electroweak symmetry breaking,
 the non-polynomial dimension-5 operator can appear
 \cite{Pierce:2006dh,Bauer:2018uxu,Bauer:2017ris,Bauer:2017nlg}
 \begin{equation}
   \frac{C^{(5)}_{aZH}}{ \Lambda} \, (\partial^\mu a) \,
   \left( \phi^\dagger i D_\mu \phi + {\rm h.c.} \right )\,
   \ln ( \phi^\dagger \phi / \mu^2 ) \;,
   \label{eq7}
 \end{equation}
which can be understood by thinking of $\phi$ as a background field
and treating the heavy particle mass as a threshold for the
running of gauge couplings.
\footnote{ A possibilty of generating such an operator can be made by
a triangular loop with a neutral heavy lepton $N$ of mass TeV
running in the loop, where $N$ is the neutral component of an
$SU(2)$ doublet and the charged component $L$ is assumed to be much heavier.
In such a setup, the $H\gamma\gamma$ and $Ha\gamma$ couplings are suppressed
by the mass of $L$. Also, the $HWW$ and $HZZ$ couplings are unlikely to
receive significant contributions from the triangular loops. The current
mass limit on heavy neutral leptons is only about 600 GeV with
$|V_{\mu N}|^2 \simeq 0.1$ \cite{Grancagnolo:2024vda}.
}
Therefore, we can write an effective coupling for $a ZH$ as
\begin{equation}
   C^{\rm eff}_{aZH} = C^{(5)}_{aZH} + \frac{C_{aZH} v^2 }{2 \Lambda^2} \;.
\end{equation}
We can now see that $H\to Z a $ is only
suppressed by one power of the cutoff scale $\Lambda$ on amplitude level
while $ H\to aa $ by two powers of $\Lambda$. That is the reason why
$H\to Z a $ can be made sizable while keeping $H\to aa$
suppressed even in the case that both coefficients $C_{ah}$ and
$C^{\rm eff}_{aZH}$ are of order $O(1)$.

  Note that the operator in Eq.~(\ref{eq7}) can induce a coupling among the
    $H$-$a$-$f\bar f$ after applying the equation of motion and integration
  by parts. Such a coupling can give rise to the rare decay
    $ H \to b \bar b a$. Nevertheless, it is highly suppressed by
  $\Lambda$ and the relatively small Yukawa $m_b/v$.

  Before we end this section, we highlight existing constraints on other
  ALP-gauge couplings denoted by $g_{a\gamma\gamma}$, $g_{aZZ}$, $g_{aZ\gamma}$,
  and $g_{aWW}$. A dedicated study on the $g_{aZZ}$, $g_{aZ\gamma}$, and
  $g_{aWW}$ was performed in Ref.~\cite{Cheung:2024qge} (references therein).
  These couplings can give rise to $pp \to Z a \to (l^+ l^-) (\gamma \gamma)$
  and $pp\to W a \to (l \nu) (\gamma \gamma)$,
  in which the photon pair can be resolved for larger $m_a$ but
  unresolved for smaller $m_a$. Other existing collider constraints on
  $g_{aZZ}$, $g_{aZ\gamma}$, $g_{aWW}$, and $g_{a\gamma\gamma}$  have
  been discussed in Ref.~\cite{Cheung:2024qge}.
  Also, $Z\to a \gamma$, which looks like $Z\to \gamma\gamma$ when
  $m_a$ is very small, was also searched in $Z\to \gamma\gamma$
  summarized in \cite{Bauer:2017ris}.
  On the other hand, comprehensive coverage of
  astrophysical constraints on $g_{a\gamma\gamma}$ can be found
  in {\tt https://cajohare.github.io/AxionLimits/}.

\section{ Results}
For convenience of calculations we can write the effective vertex for
$aZH$,
after the electroweak symmetry breaking, as 
 \begin{equation}
   {\cal L}_{aZH} = \frac{ C^{\rm eff}_{aZH} }{\Lambda} \frac{g v}{c_w}
  \left( \partial^\mu a \right)\,  Z_\mu \, H
 \end{equation}
  which implies the Feynman rule in Fig.~\ref{fey}. Here $v \simeq 246$ GeV
  and $c_w$ is the cosine of the Weinberg angle.

\begin{figure}[th!]
  \includegraphics[width=5.5in]{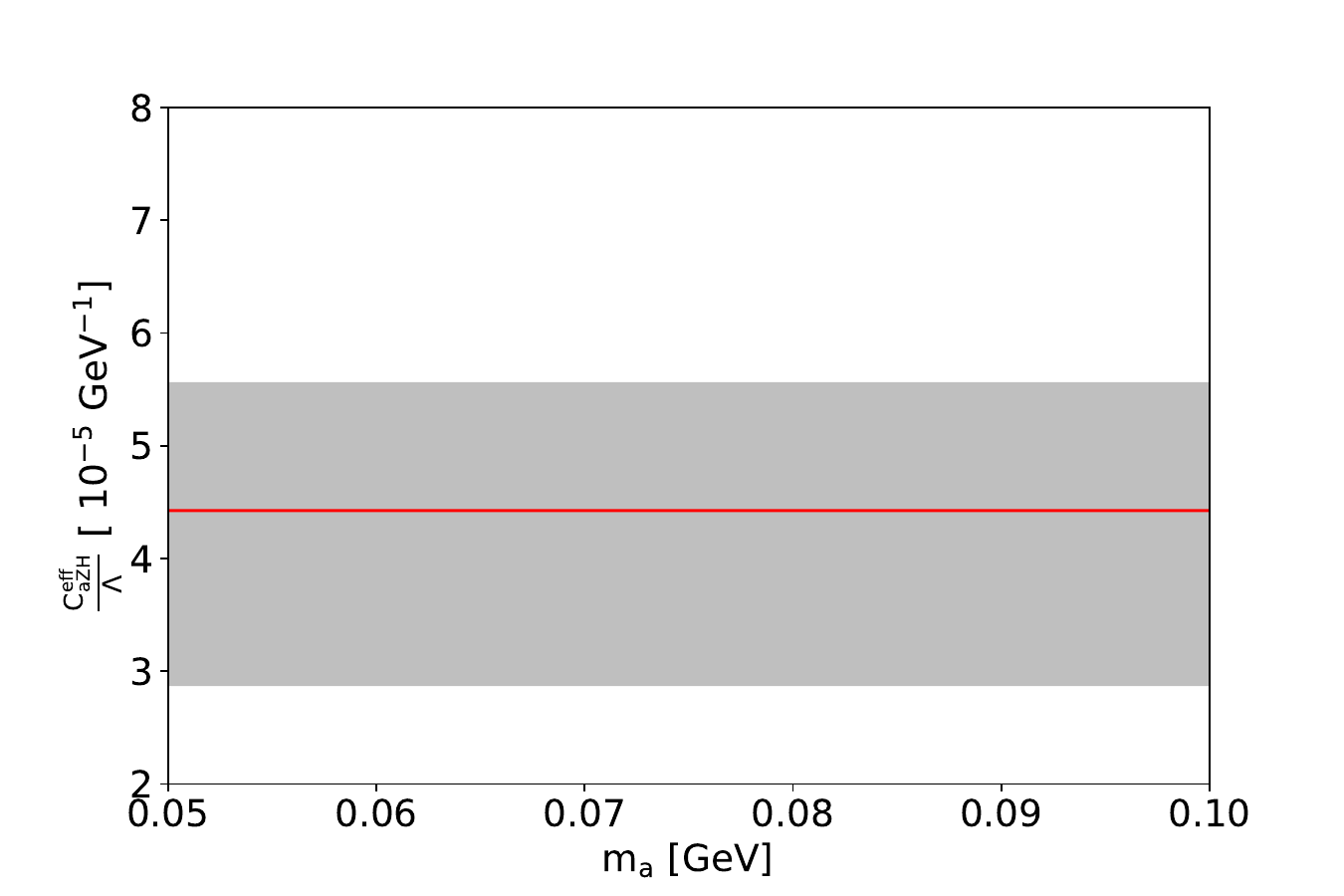}
  \caption{\small \label{contour}
    The fitted values for $C^{\rm eff}_{aZH} / \Lambda$ versus $m_a$ for $m_a
    =0.05 - 0.1$ GeV.
    The red line and the band show the central value and $1\sigma$
     uncertainty in $C^{\rm eff}_{aZH} /\Lambda = (4.4 \pm 1.1) \times 10^{-5}
     \;{\rm GeV}^{-1}$  corresponding to
     $B(H\to Z \gamma ) = ( 1.9 \times 1.1 ) \times 10^{-3}$.
  }
\end{figure}

We can calculate the partial width of $ H \to Z a$ and $H\to a a $
\cite{Bauer:2018uxu}
\begin{eqnarray}
  \Gamma (H \to Z a ) &=&
    \frac{m^3_H}{16 \pi} \left( \frac{ C^{\rm eff}_{aZH}} {\Lambda} \right )^2 \,
    \lambda^{3/2} ( x_Z, x_a ) \\
    \Gamma (H \to a  a ) &=&
      \frac{m^3_H v^2 }{32 \pi} \left( \frac{ C_{aH}} {\Lambda^2} \right )^2 \,
    ( 1 - 2 x_a )^2  \sqrt{ 1 - 4 x_a} \;,
\end{eqnarray}
where $x_i = m_i^2 / m_H^2 \;(i=a, Z)$ and $\lambda(x,y) = (1-x-y)^2 - 4 xy$.
Including the new contribution of $\Gamma( H\to Z a)$
the branching ratio of $H\to Z a $ is given by
\begin{equation}
  B( H\to Z a  ) = \frac{ \Gamma (H \to Z a) }
  { \Gamma (H \to Z a ) + \Gamma_{\rm sm} (m_H = 125 \,{\rm GeV} )  } \;.
\end{equation}
where $\Gamma_{\rm sm} (m_H = 125\;{\rm GeV})$ is taken to be
$4.088 \times 10^{-3}$ GeV for $m_H = 125$ GeV
\footnote{
It is available at
{\tt https://twiki.cern.ch/twiki/bin/view/LHCPhysics/CERNYellowReportPageBR}.
}.

Requiring the branching ratio to be $(1.9 \pm 1.1)\times 10^{-3}$ as in
Eq.~(\ref{1.9} ), we obtain the results as shown in Fig.~\ref{contour}.
We found that
\begin{equation}
  \frac{C^{\rm eff}_{aZH}}{ \Lambda}
  = ( 4.4\;^{+1.1}_{-1.6} ) \times 10^{-5}\;{\rm GeV}^{-1}
\end{equation}
where the upper and lower limits correspond to the $1\sigma$ of
$B(H\to Za) = (1.9 \pm 1.1)\times 10^{-3}$. If the coefficient
$C^{\rm eff}_{aZH} \sim  O(1)$ the corresponding cutoff scale is
$\Lambda = 22.6$ TeV.

\begin{figure}[th!]
  \centering
  \includegraphics[width=5.5in]{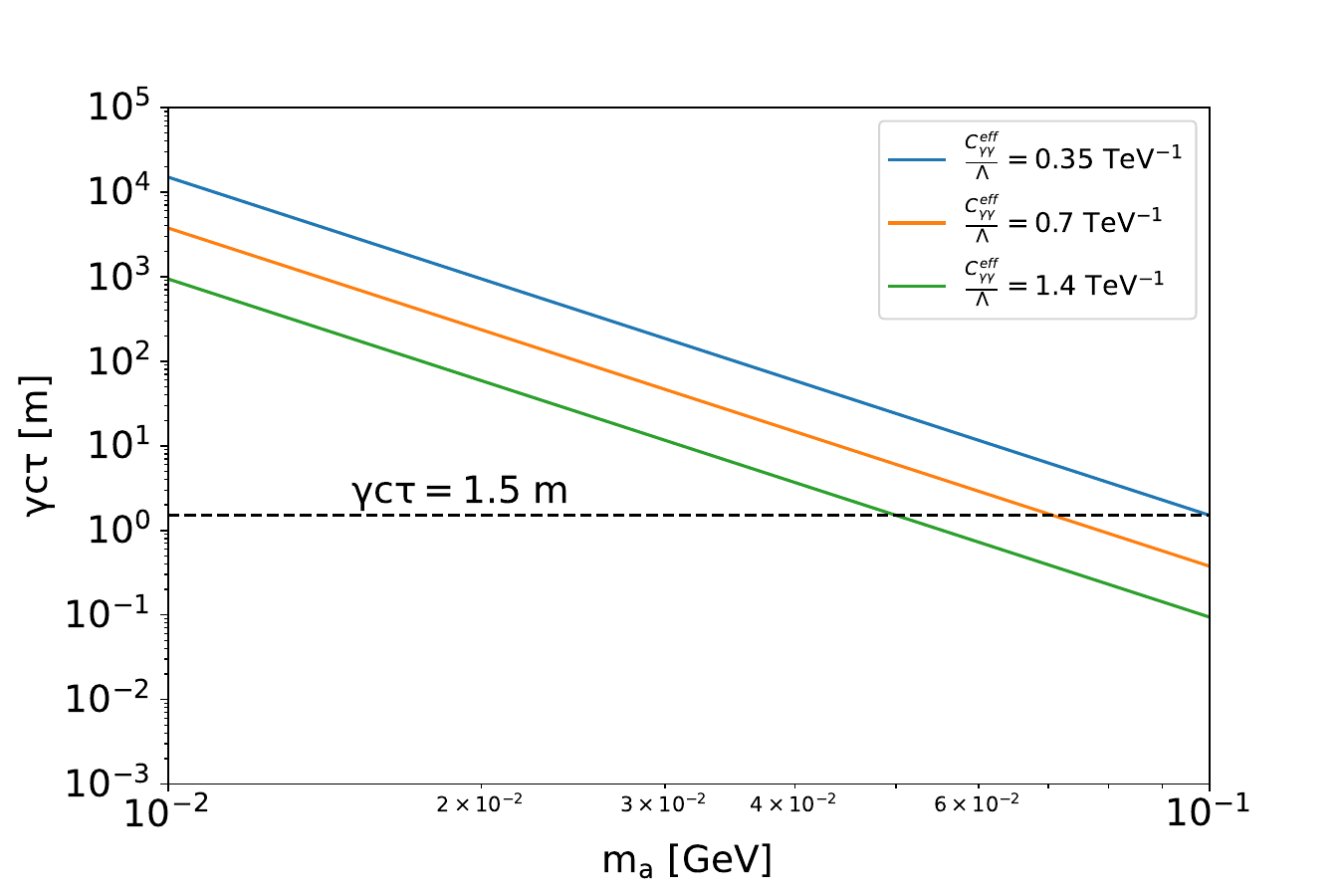}
  \caption{\small \label{decaylength}
    Decay length $\gamma c \tau$ versus the mass $m_a$ of the ALP.
    Values of $C^{\rm eff}_{\gamma\gamma} / \Lambda = 0.35,\, 0.7,\, 1.4 \;
    {\rm TeV}^{-1}$ are used. A dashed  horizontal line of 1.5 m is also
    shown.
  }
\end{figure}

  It is true that the result of $C^{\rm eff}_{aZH} /\Lambda$
   corresponds to the mass scale of $\Lambda = 22.6$ TeV
   with $O(1)$ coefficient. If it is the case, these heavy particles
   would certainly be out of reach at the LHC. On the other hand,
   if we take the coefficient to be $O(0.1) \approx e^2$
   (as in the definition of
   $e^2 (C_{\gamma\gamma}/\Lambda) a F_{\mu\nu} \tilde{F}^{\mu\nu}$),
   the scale $\Lambda$ would then be around 2 TeV, which may be
   readily available at the LHC. Indeed, the current mass limit on
   heavy vector-like quarks is about $O(1) - 1.6$ TeV
   depending on the search channels (for a recent review see
   \cite{CMS:2024bni}), and the mass limit on heavy neutral leptons
   is about 600 GeV with $|V_{\mu N}|^2 \simeq 0.1$
   \cite{Grancagnolo:2024vda}.

\begin{figure}[th!]
  \centering
  \includegraphics[width=6in]{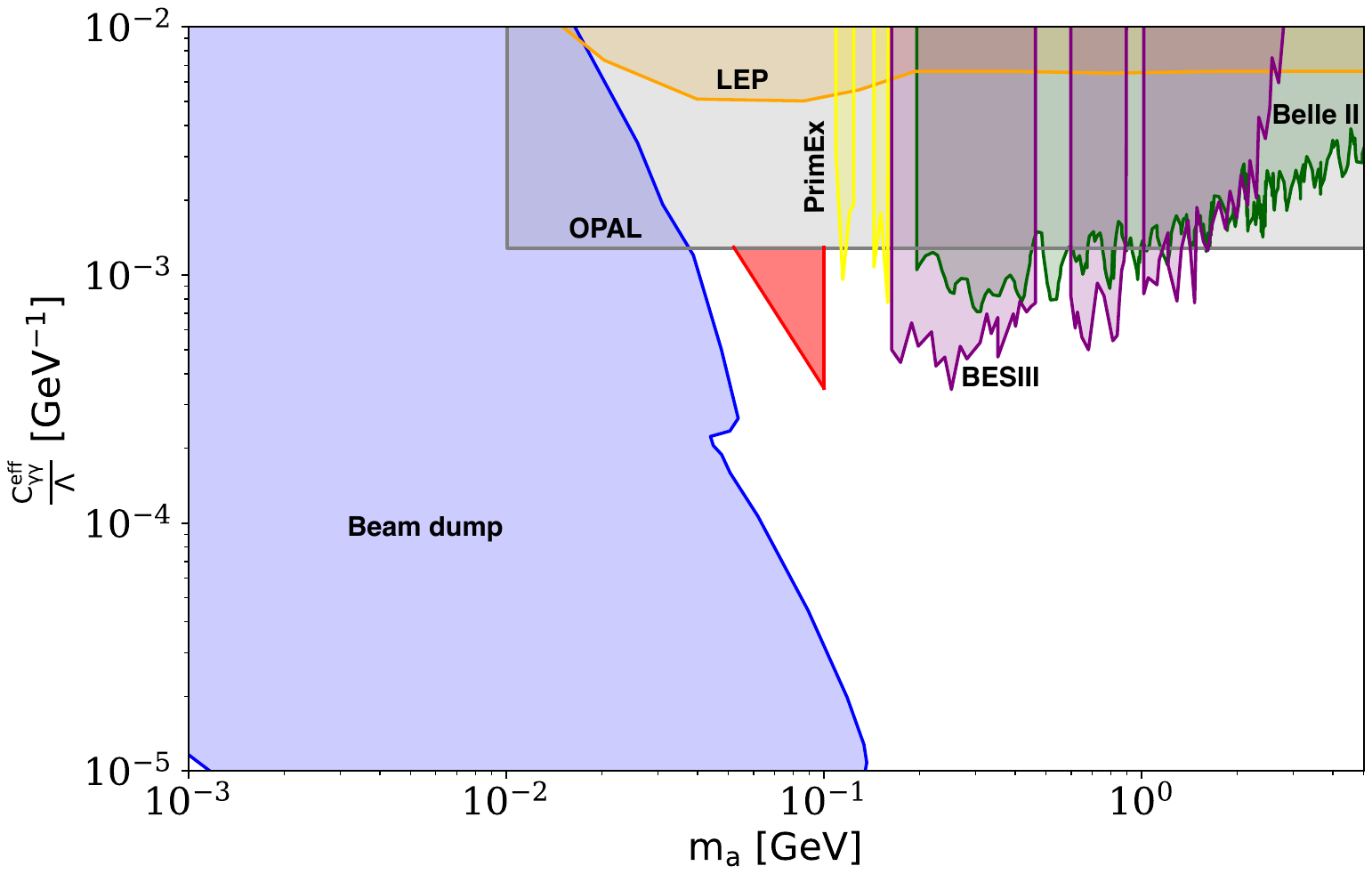}
  \caption{\small \label{cgamma}
    Parameter space (shaded in red)
    in $C^{\rm eff}_{\gamma\gamma} / \Lambda$ versus $m_a$
    that can allow the ALP to decay before reaching the ECAL (i.e.
    $\gamma c \tau \le 1.5$ m) and
    consistent with all existing constraints in the mass range of
    $10^{-3} - 5$ GeV,
  including beam dump~\cite{CHARM:1985anb,Riordan:1987aw,Dolan:2017osp,Dobrich:2019dxc,NA64:2020qwq},
      OPAL~\cite{Knapen:2016moh}, LEP~\cite{Jaeckel:2015jla},
      Belle II~\cite{Belle-II:2020jti}, BES III~\cite{BESIII:2022rzz},
      and PrimEx~\cite{PrimEx:2010fvg}
      (data extracted from the GitHub page~\cite{AxionLimits}).
      Note that the mass range of the fitted parameter space is
        $0.05\,{\rm GeV} \le m_a \le 0.1\,{\rm GeV}$.
  }
\end{figure}

Let us turn to the requirement on the coupling
$C^{\rm eff}_{\gamma\gamma} / \Lambda$,
which controls the decay length $\gamma c \tau$ of the ALP, where
$\gamma = E_a / m_a \; (E_a \approx m_H/2) $ is
the Lorentz boost factor of the ALP and the decay time
$\tau$ in the rest frame is given by
\begin{equation}
  \tau = \frac{1}{\Gamma_a}\,, \qquad
\Gamma_a = 4 \pi \alpha^2 m_a^3 \left( \frac{C^{\rm eff}_{\gamma\gamma}}{\Lambda}
\right)^2 \;,
\end{equation}
where $\Gamma_a$ is the total decay width of the ALP assuming it only decays
into diphoton.
We show the decay length of the ALP versus $m_a$ for a few values of
$C^{\rm eff}_{\gamma\gamma}/ \Lambda$ in Fig.~\ref{decaylength}. Taking
the input values of $E_a = m_H/2 = 62.5$ GeV, the requirement of the
$\gamma c \tau \le 1.5 \,{\rm m}$ gives
\begin{equation}
  \frac{ C^{\rm eff}_{\gamma\gamma} }{ \Lambda} \ge 0.35 \;{\rm TeV}^{-1} \,
  \left( \frac{0.1\,{\rm GeV}}{m_a} \right)^2 \;.
\end{equation}
Therefore, at $m_a = 0.1 \, (0.05)$ GeV the coupling
$C^{\rm eff}_{\gamma\gamma} / \Lambda > 0.35\,(1.4) \, {\rm TeV}^{-1}$.
We show in Fig.~\ref{cgamma} the region of parameter space in
$(m_a,\,C^{\rm eff}_{\gamma\gamma} / \Lambda)$ that can allow the ALP to
decay before reaching the ECAL and consistent with all existing
constraints.
  Note that the lower mass limit $m_a = 0.05$ GeV is due to the
  existing constraints (see Fig.~\ref{cgamma}),
  while the upper limit $m_a=0.1$ GeV
  came from the limitation of the shower-shape analysis \cite{ATLAS:2012soa}.

Such a scenario using a light axion with the diphoton decay,
which mimics a single photon, to explain the excess in
$H \to Z \gamma$ can be tested at the $Z$ resonance (Tera-$Z$ -- $10^{12} \,Z$
bosons) of the Future Circular Colliders \cite{FCC:2018byv} and
CEPC \cite{CEPCStudyGroup:2018ghi}.
Via the same coupling $C^{\rm eff}_{aZH} / \Lambda $ the $Z$ boson
can decay via an off-shell Higgs boson
\[
Z \to a H^* \to a (b \bar b) \;,
\]
in which the most dominant mode of the virtual Higgs boson is considered.
The final state consists of a $b\bar b$ pair plus a diphoton, which
appears as a single photon. Nevertheless, the branching ratio is only
$10^{-12}$, which barely affords a few events at the Tera-$Z$ option.

Another possible test of the scenario is the production process
$pp \to Z^* \to a H $ at the LHC via the same coupling
$C^{\rm eff}_{aZH}/ {\Lambda}$. However,
the cross section turns out to be negligible,
of order $10^{-6}$ fb only,
which corresponds to far less than 1 event
  for 3000 fb$^{-1}$ luminosity of the entire running of High-Luminosity LHC
  (HL-LHC).

\section{Conclusions}

The excess observed in the rare decay of the Higgs boson
into a $Z$ boson and a photon can be interpreted as the Higgs decay into
a $Z$ boson and a light axion. The light axion then decays into a pair
of collimated photons such that the ECAL cannot resolve.  Such a scenario
requires a coupling between $aZH$ with strength
$C^{\rm eff}_{aZH} /\Lambda \sim 4 \times 10^{-5}\;{\rm GeV}^{-1}$.
Furthermore, the axion is required to decay before reaching the ECAL,
which implies the effective $C^{\rm eff}_{\gamma\gamma} / \Lambda \ge
0.35 \,{\rm TeV}^{-1} \, (0.1\,{\rm GeV}/m_a)^2$.
Such a $aZH$ coupling may be tested via
$Z \to a H^* \to a (b\bar b)$ at the Tera-$Z$ option
of the FCC and CEPC, but, however, it has a branching ratio of
only $10^{-12}$.

Note that the cutoff scale defined in $C^{\rm eff}_{aZH} / \Lambda$ is
  about $22.6$ TeV with $O(1)$ coefficient.
  On the hand, the requirement of the decay length of $\gamma c\tau \le
  1.5$ m  needs $C^{\rm eff}_{\gamma\gamma} /\Lambda = 0.35 \, {\rm TeV}^{-1}
  (0.1 \,{\rm GeV}/m_a)^2$.
  Note that there is a factor $e^2 \sim 0.1$ in front of
  $C_{\gamma\gamma} /\Lambda$ in Eq.~(\ref{eq:gamma}). Therefore, if we take
  out this factor $e^2$, the corresponding $\Lambda$ with $O(1)$ coefficient
  would become $28$ TeV for $m_a = 0.1$ GeV and $7$ TeV for $m_a=0.05$ GeV,
  such that these two sets of $\Lambda$'s are of similar order.
  Nevertheless, these two values are purely phenomenological.

  A comment on the constraints from flavor-changing processes is in
order here.  The ALP with mass $m_a = 0.05 - 0.1$ GeV may subject to
constraints from flavor-changing processes such as $K \to \pi \nu \bar
\nu$ and $K \to \pi \mu^+ \mu^-$ and the corresponding ones for $B\to
K$. It was shown in Ref.~\cite{Bauer:2021mvw} that $C_{WW}$ coupling can
induce ALP flavor-changing coupling at one-loop order such that the
constraints on $C^{\rm eff}_{\gamma\gamma}/\Lambda$ strongly restrict
our fitted region of Fig.4. On the other hand, if only $C_{BB}$
coupling exists in the UV scale, the flavor-changing couplings
involving the ALP can only be generated at two loops \cite{Bauer:2021mvw},
and are therefore highly suppressed. As shown in the right panel of Fig. 22
of Ref.~\cite{Bauer:2021mvw}, the flavor constraints are rather weak in this
case and our fitted parameter space is valid.

{\it Acknowledgment}.
The work was supported by the MoST of Taiwan under Grants 
MOST-110-2112-M-007-017-MY3.

\bibliographystyle{JHEP}
\bibliography{paper}

\end{document}